\documentclass[aps,twocolumn]{revtex4-1}

\usepackage{amsmath,MnSymbol}
\usepackage{graphicx}

\newcommand{\lambert}{\ensuremath{{\cal W}}}

\newcommand{\sopa}{SUPA, School of Physics and Astronomy, University of Edinburgh, Edinburgh EH9 3JZ, UK}
\newcommand{\ppls}{Language Evolution and Computation Research Unit, School of Philosophy, Psychology and Language Sciences, University of Edinburgh, Edinburgh EH8 9LL, UK}
\newcommand{\tum}{Fakult\"at f\"ur Physik, Technische Universit\"at M\"unchen, James-Franck-Str.~1, 85748 Garching, Germany}

\begin{document}

%: Frontmatter
\title{Stochastic dynamics of lexicon learning in an uncertain and nonuniform world}
\date{May 31, 2013}

\author{Rainer Reisenauer}
\affiliation{\tum}
\affiliation{\sopa}

\author{Kenny Smith}
\affiliation{\ppls}

\author{Richard A.\ Blythe}
\affiliation{\sopa}

\begin{abstract}
We study the time taken by a language learner to correctly identify the meaning of all words in a lexicon under conditions where many plausible meanings can be inferred whenever a word is uttered. We show that the most basic form of \emph{cross-situational learning}---whereby information from multiple episodes is combined to eliminate incorrect meanings---can perform badly when words are learned independently and meanings are drawn from a nonuniform distribution. If learners further assume that no two words share a common meaning, we find a phase transition between a \emph{maximally efficient} learning regime, where the learning time is reduced to the shortest it can possibly be, and a \emph{partially-efficient} regime where incorrect candidate meanings for words persist at late times. We obtain exact results for the word-learning process through an equivalence to a statistical mechanical problem of enumerating loops in the space of word-meaning mappings.
\end{abstract}

\maketitle

%: Intro

On average, children learn ten words a day, thereby amassing a lexicon of 60,000 words by adulthood \cite{Bloom00}. This speed of learning is remarkable given that every time a speaker says a word, a hearer cannot be certain of its intended meaning \cite{Quine60}. Our aim is to identify which of the many proposed mechanisms for eliminating uncertainty can actually deliver such rapid word learning. In this work, we pursue this aim in the long tradition of applying quantitative methods from statistical mechanics to problems in learning \cite{Watkin93,Hopfield82,Amit85a,Amit85b} and communication \cite{Sourlas89,Kabashima00,Nishimori01}.

Empirical research suggests that two basic types of learning mechanism are involved in word learning. First, a learner can apply various \emph{heuristics}---e.g., attention to gaze direction \cite{Baldwin91} or prior experience of language structure \cite{*[{See e.g.~}] [{ for a review.}] Bloom98}---at the moment a word is produced to hypothesize a set of plausible meanings. However, these heuristics may leave some residual uncertainty as to a word's intended meaning in a single instance of use. If the heuristics are weak, the set of candidate meanings could be very large. This residual uncertainty can be eliminated by comparing separate instances of a word's use: if only one meaning is plausible across all such instances, it is a very strong candidate for the word's intended meaning. This second mechanism is referred to as \emph{cross-situational learning} \cite{Pinker89,Siskind96}. Formally, it can be couched as a process whereby associations between words and meanings are strengthened when they co-occur \cite{Siskind96,Vogt04,Tilles12a,Yu12}, as in neural network models for learning \cite{Watkin93,Hopfield82,Amit85a,Amit85b,Pulvermuller99}. It can also be viewed as an error-correction process \cite{Sourlas89,Kabashima00,Nishimori01} where a target set of associations is reconstructed from noisy data.

There is little consensus as to which word-learning mechanisms are most important in a real-world setting \cite{Gleitman90,Pinker94,Yu07,Frank09,Medina11}. In part this is because word-learning experiments (e.g.~\cite{Yu07,Smith08,Smith11}) are necessarily confined to small lexicons. A major question is whether strategies observed in experiments allow realistically large lexicons to be learned rapidly: this can be fruitfully addressed through stochastic dynamical models of word learning \cite{SSBV06,BSS10,Tilles12a,Tilles12b}. In these models, a key control parameter is the \emph{context size}: the number of plausible, but unintended, meanings that typically accompany a single word's true meaning. Even when contexts are large, the rapid rate of learning seen in children is reproduced in models where words are learned independently by cross-situational learning \cite{SSBV06,BSS10,Tilles12a,Tilles12b}. This suggests that powerful heuristics, capable of filtering out large numbers of spurious meanings, are not required. However, a recent simulation study \cite{Vogt12} shows that this conclusion relies on the assumption that these unintended meanings are uniformly distributed. In the more realistic scenario where different meanings are inferred with different probabilities, word learning rates can decrease dramatically as context sizes increase. Powerful heuristics may be necessary after all.

One heuristic, of great interest to empiricists (e.g.~\cite{Markman88,golinkoff_94_young,halberda_03_development,Markman03}) and modelers (e.g.~\cite{Frank07,BSS10,Tilles12a,Tilles12b,Vogt12}), is a \emph{mutual exclusivity constraint} \cite{Markman88}. Here, a learner assumes that no two words may have the same meaning.  This generates nontrivial interactions between words which makes analysis of the corresponding models difficult.  For example, if one begins with a master equation, as in \cite{SSBV06,BSS10,Tilles12a}, the expressions become unwieldy to write down, let alone solve. Here, we adopt a fundamentally different approach which entails identifying the criteria that must be satisfied for a lexicon to be learned. This allows existing results for the simple case of independently-learned words and uniform meaning distributions \cite{BSS10} to be generalized to arbitrary meaning distributions \emph{and} exactly solves the interacting problem  to boot. Our main result is that mutual exclusivity induces a dynamical phase transition at a critical context size, below which the lexicon is learned at the \emph{fastest} possible rate (i.e., the time needed to encounter each word once). As far as we are aware, the ability of a single heuristic to deliver such fast learning has not been anticipated in earlier work.

%: Model definition

\begin{figure}[tb]
\includegraphics[width=\linewidth]{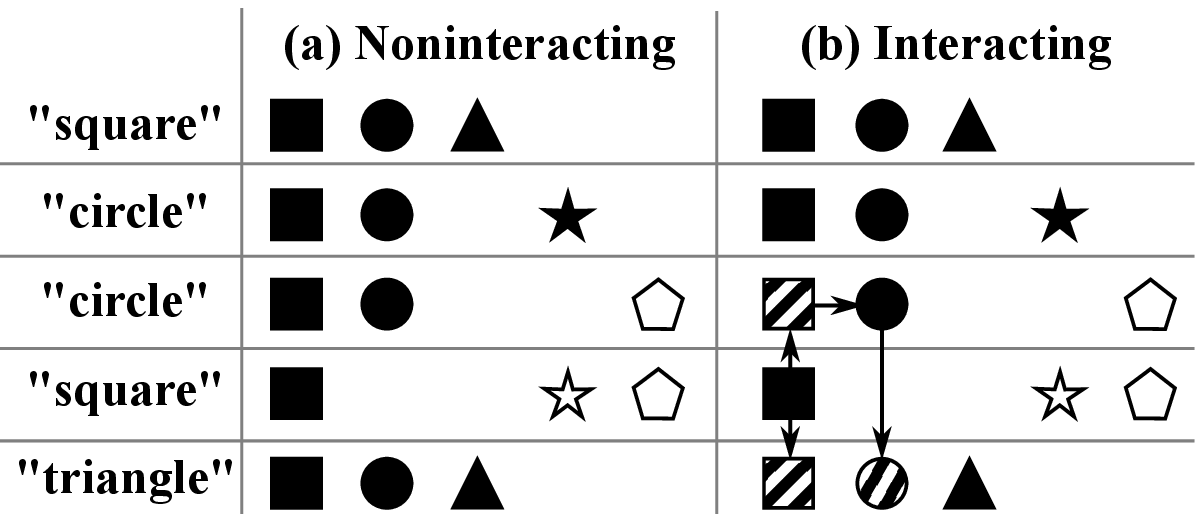} % Word equivalence = 84
\caption{\label{XSL} Acquisition of a three-word lexicon. Solid shapes are meanings that have appeared in every episode alongside a word; open shapes are therefore excluded as candidate meanings. (a) In the noninteracting case, only the meaning of the word `'square'' is learned. (b) In the interacting case, mutual exclusivity further removes meanings (shown hatched) of learned words, both prospectively and retrospectively (shown by arrows). All three words are learned in this example.}
\end{figure}

We begin by defining our model for lexicon learning.  The lexicon comprises $W$ words, and each word $i$ is uttered as a Poisson process with rate $\phi_i$.  In all cases, we take words to be produced according to the Zipf distribution, $\phi_i = 1/(\mu i)$, that applies for the ${\sim}10^4$ most frequent words in English \cite{Zipf49,Cancho01,Petersen12}. Here, $\mu=\sum_{i=1}^{W} (1/i)$ so that one word appears on average per unit time. Each time a word $i$ is presented, the intended \emph{target} meaning is assumed always to be inferred by the learner by applying some heuristics. At the same time, a set of non-target \emph{confounding} meanings, called the \emph{context}, is also inferred.

In the purest version of cross-situational learning \cite{Siskind96,BSS10}, a learner assumes that all meanings that have appeared every time a word has been uttered are plausible candidate meanings for that word.  The word becomes \emph{learned} when the target is the only meaning to have appeared in each episode. In the \emph{noninteracting} case, each word is learned independently---see Fig.~\ref{XSL}a. In the \emph{interacting} case, mutual exclusivity acts to further exclude the meanings of learned words as candidates for other words. We take this exclusion to occur at the instant a word is learned, which means a single learning event may trigger an avalanche of other learning events by repeated application of mutual exclusivity. An example of this nontrivial effect that is hard to handle within standard approaches \cite{BSS10,Tilles12a} is shown in Fig.~\ref{XSL}b. Here, learning ``square'' causes ``circle'' to be learned at the same time.

%: Nonnteracting problem

We consider the noninteracting case first both to introduce our more powerful analytical approach and to pinpoint the origin of the catastrophic increase in learning times noted in \cite{Vogt12}.  Two conditions must be satisfied for the lexicon to be learned by a given time: (C1) all words must have been exposed at least once; and (C2) no confounding meaning may have appeared in every episode that any given word was uttered.  To express these conditions mathematically, we introduce two stochastic indicator variables.  We take $E_i(t)=1$ if word $i$ has been uttered before time $t$, and zero otherwise; and $A_{i,j}(t)=1$ if confounding meaning $j$ has appeared in every context alongside word $i$ up to time $t$ (or if word $i$ has never been presented), and zero otherwise.  Conditions (C1) and (C2) then imply that the probability that the lexicon has been learned by time $t$ is
\begin{equation}
\label{Lni}
L(t) = \Big\langle \prod_{i} E_i(t) \prod_{j\ne i} [1 - A_{i,j}(t)] \Big\rangle =  \Big\langle \prod_{i\ne j} [1 - A_{i,j}(t)] \Big\rangle
\end{equation}
where the angle brackets denote an average over all sequences of episodes that may occur up to time $t$.  The second equality holds because $A_{i,j}(t) = 1 \forall j\ne i$ if $E_i(t)=0$.

This expression is valid for any distribution over contexts. For brevity, we consider a single, highly illustrative construction that we call \emph{resampled Zipf} (RZ). It is based on the idea that meaning frequencies should follow a similar distribution to word forms \cite{Vogt12}. It works by associating an ordered set, ${\cal M}_i$, of $M$ confounding meanings with each word $i$. The $k^{\rm th}$ meaning in each set has an \emph{a priori} statistical weight $1/k$.  Whenever word $i$ appears, meanings are repeatedly sampled from ${\cal M}_i$ with their \textit{a priori} weights, and added to the context if they are not already present until a context of $C$ distinct meanings has been constructed. When words are learned independently, the learning time depends only on $M$, $W$ and $C$, and not on which meanings are present in any given set ${\cal M}_i$ \cite{BSS10}.

We seek the time, $t^\ast$, at which the lexicon is learned with some high probability $1-\epsilon$.  In the RZ model, each context is an independent sample from a fixed distribution. Hence, the correlation functions $\langle A_{i_1,j_1} A_{i_2,j_2} \cdots \rangle$ in (\ref{Lni}) all decay exponentially in time.  To find $t^\ast$ to good accuracy in the small-$\epsilon$ limit, only the slowest decay mode for each word $i$ is needed.  Higher-order correlation functions depend on many meanings co-occurring, and so decay more rapidly than lower-order correlation functions. As shown in Appendix~\ref{app:a}, we find that at late times (\ref{Lni}) is well approximated by
\begin{equation}
\label{Lnia}
L(t) \sim \prod_{i} \big[ 1 - {\rm e}^{-\phi_i (1-a_i^\ast) t} \big]
\end{equation}
where $a_i^\ast$ is the fraction of episodes in which word $i$'s most frequent confounder appears alongside the target. This expression generalizes results for independently-learned words  \cite{SSBV06,BSS10,Tilles12a} from uniform to \emph{arbitrary} nonuniform confounder distributions.

The RZ model has the further simplification that $a_i^\ast$ has a common value, $a^\ast$, for all words $i$. Then, it is known from previous calculations \cite{BSS10} for Zipf-distributed word frequencies that the learning time is
\begin{equation}
\label{tast1}
t^\ast \sim \frac{\mu W}{1-a^\ast} \lambert_0\left( \frac{W}{-\ln(1-\epsilon)} \right)
\end{equation}
where $\lambert_0(z)$ is the principal branch of the Lambert W function \cite{Corless96}. For large argument, this function behaves as a logarithm.  

\begin{figure}[tb]
\includegraphics[width=\linewidth]{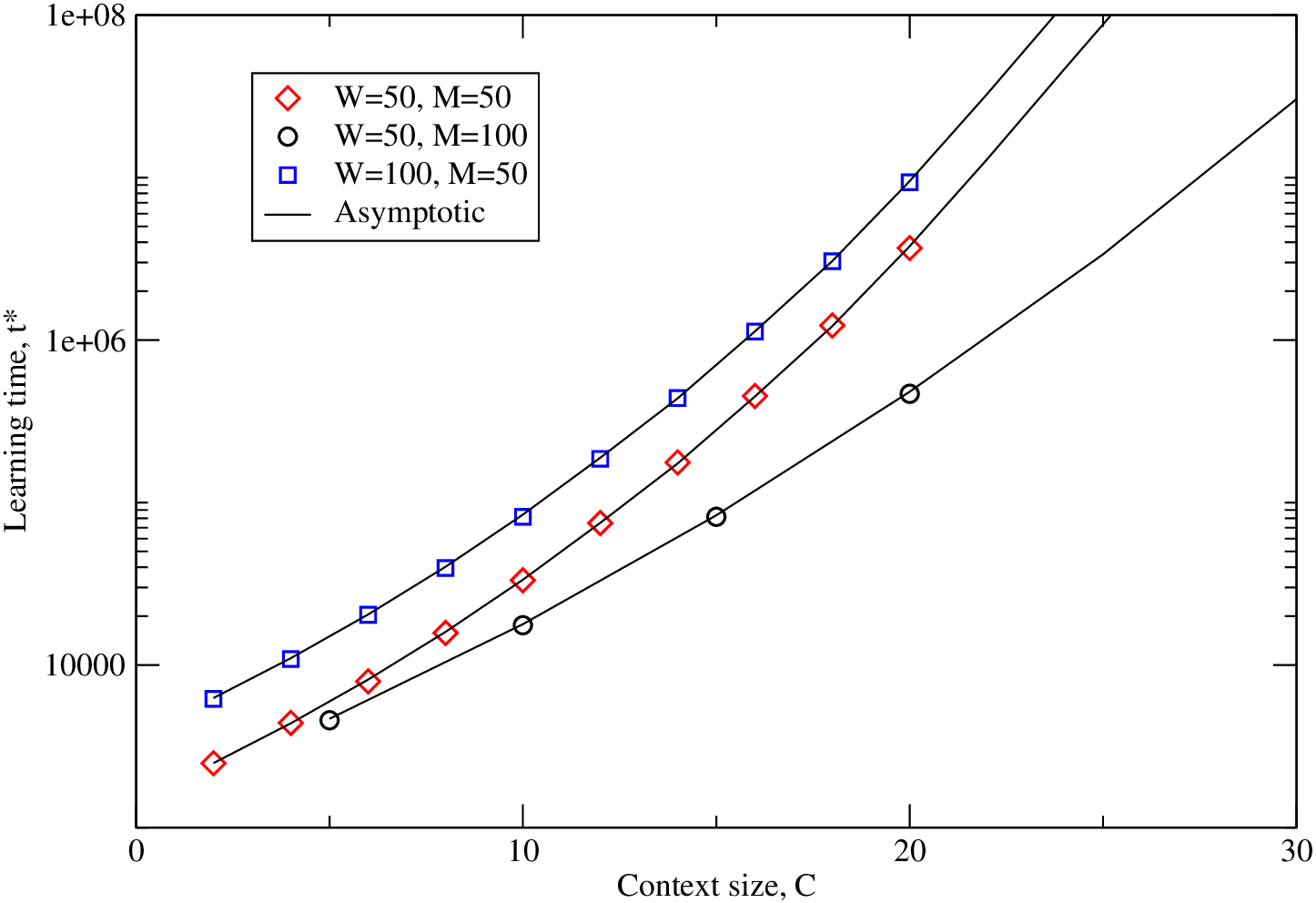} % Word equivalence = 123
\caption{\label{RZ1} Time to learn a lexicon of $W$ words independently to a residual probability $\epsilon=0.01$ with $C$ of $M$ confounders present in each episode. Points: data from Monte Carlo simulations (over $10,000$ sampled lexicons in each case). Lines: the analytical result, Eq.~(\ref{tast1}).}
\end{figure}

In Fig.~\ref{RZ1}, we compare the analytical result (\ref{tast1}) with learning times obtained from direct Monte Carlo simulations, conducted as detailed in \cite{BSS10}.  The only complication is that we unfortunately have no analytic expression for $a^\ast$ arising from the RZ procedure.  We therefore obtain the frequency of the most common confounder for given $C$ and $M$ from independent Monte Carlo samples.  The agreement between (\ref{tast1}) and simulation is very good.

Fig.~\ref{RZ1} also shows that the learning time increases super-exponentially with the context size.  We have found that the probability the $k^{\rm th}$ most confounder appears in a context of size $C$ fits the form $p_k \approx 1-(1-w_k)^{C{\rm e}^{\lambda C}}$ where $w_k$ is the \emph{a priori} probability and $\lambda$ is a fitting parameter that depends on $M$ and $k$. As noted by Vogt \cite{Vogt12}, the repeated sampling without replacement implies that $p_k \ge 1-(1-w_k)^C$. Our analysis further reveals that the learning time is \emph{entirely} determined by the frequency of the most common confounder, $a^\ast$ through (\ref{tast1}). We note that this is true even when other confounders have comparable appearance frequencies ($C \le 5$).

%: Interacting problem

We now turn to the case where the mutual exclusivity constraint serves to exclude the meanings of learned words as possible meanings for other words.  In this case, it is important to distinguish between \emph{labeled} and \emph{unlabeled} meanings: an unlabeled meaning is not the target meaning of any word in the lexicon, and hence cannot be excluded using the mutual exclusivity constraint. To generalize Eq.~(\ref{Lni}) to this problem, we must identify the conditions for the lexicon to be learned.  Condition (C1) still applies: each word must be uttered at least once for a learner to be able to learn it.  Condition (C2) now applies only to unlabeled confounding meanings: these can only be excluded if they fail to appear in a context, as before.  When these two conditions are satisfied, there is a third---necessary and sufficient---condition for the lexicon to be learned that takes into account all the interactions and avalanches generated by the mutual exclusivity constraint.  This is condition (C3): no \emph{candidate loops} exist at time $t$.  A candidate loop, $\ell=(i_1, i_2, \ldots, i_n)$, is a subset of distinct, labeled meanings whereby each meaning $i_{k}$ has appeared alongside the word associated with meaning $i_{k-1}$ (or $i_{n}$ if $k=1$) every time it has been uttered.  Inspection of Fig.~\ref{XSL}b shows that the one candidate loop (\raisebox{1pt}{$\scriptstyle\blacksquare$},$\bullet$) that exists after the third episode is destroyed in the fourth. Then, in the fifth episode, the final word appears, and since no unlabeled meaning is a candidate for any word, the entire three-word lexicon is learned.

To see why condition (C3) is necessary and sufficient in general when (C1) and (C2) hold, we first show that a candidate loop must exist if the lexicon has not been learned. Suppose word $i_1$ has not been learned. Then, at least one meaning, $i_2$, must confound word $i_1$. Word $i_2$ must also not have been learned, otherwise meaning $i_2$ would not confound word $i_1$. Hence, word $i_2$ must be confounded by a meaning, $i_3$, and so on. As there is a finite set of words, this sequence of meanings must eventually form a loop.

We now show the lexicon cannot have been learned if a candidate loop exists by first assuming that it \emph{has} been learned under these conditions. Then, if word $i_1$ was learned at time $t$, word $i_2$ must have been learned before time $t$ for mutual exclusivity to act (even if words $i_1$ and $i_2$ are learned as part of the same avalanche). Iterating this argument around the loop, one finds that word $i_1$ can only have become learned at time $t$ if it had already been learned at some earlier time. This contradiction therefore implies that the absence of candidate loops and a learned lexicon are equivalent.

We again use indicator variables to translate conditions (C1)--(C3) into an exact expression for the learning probability.  Introducing $C_{\ell}(t) = A_{i_1,i_2}(t) A_{i_2,i_3}(t) \cdots A_{i_n,i_1}(t)$ that equals $1$ if the loop $\ell$ persists at time $t$, we have
\begin{equation}
\label{lme}
L(t) = \left\langle \prod_{i=1}^{W} E_i(t) \prod_{j>W} \left[ 1 - A_{i,j}(t) \right] \prod_{\ell} \left[ 1 - C_{\ell}(t) \right] \right\rangle \;,
\end{equation}
again valid for any distribution of confounding meanings. Here, meanings $1$ to $W$ correspond to words $1$ to $M$, and so meanings with an index $j>W$ are unlabeled. The product over $\ell$ is over all possible candidate loops.  This expression has the remarkable property that it is expressed concisely in terms of the word and confounder appearance frequencies alone: the avalanche dynamics triggered by mutual exclusivity do not enter explicitly.  This property, reminiscent of the avalanche dynamics of Abelian sandpile models \cite{Dhar90}, reduces analysis of the learning probability to the statistical mechanical problem of enumerating candidate loops.

\begin{figure}[tb]
\includegraphics[width=\linewidth]{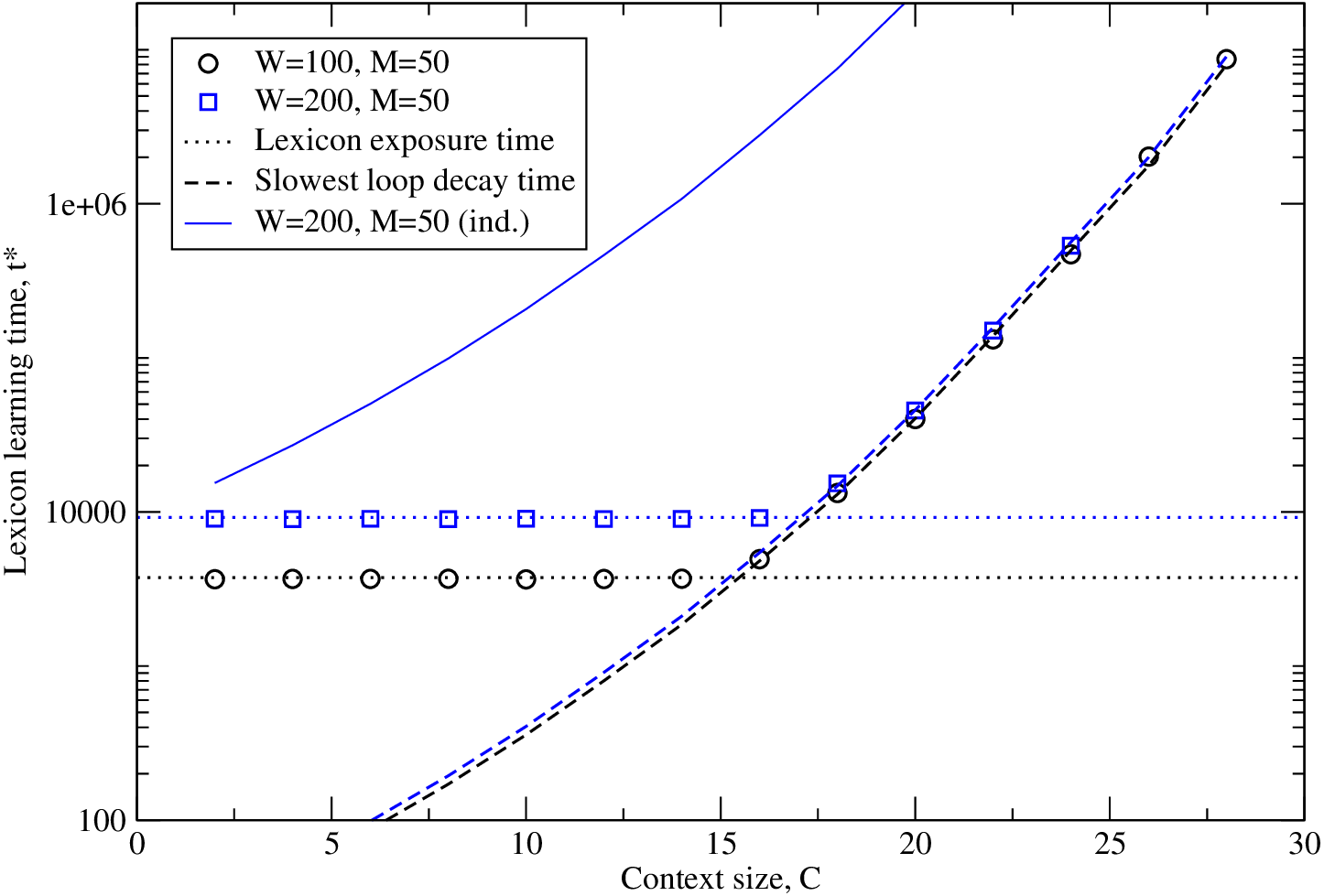} % Word equivalence = 122
\caption{\label{RZL} As Fig.~\ref{RZ1} but with the mutual exclusivity constraint. Points: data from Monte Carlo simulations (100,000 lexicons for $C \le 20$, at least 2,500 lexicons for larger $C$). Dotted lines: time for the entire lexicon to have been exposed with residual probability $\epsilon=0.01$. Dashed lines: time for the slowest decaying candidate loop to remain with probability $\epsilon$. Solid line: time to learn lexicon independently, Eq.~(\ref{tast1}), for comparison.}
\end{figure}

In the interacting problem, the structure of each candidate set ${\cal M}_i$ is important, as this determines which words interact. We consider a model which has no unlabeled meanings and where each set ${\cal M}_i$ is a sample of $M$ non-target meanings obtained via the RZ prescription. Then, in each episode, $C$ meanings are drawn from the relevant candidate set using RZ again, but with an \emph{a priori} weight $1/k$ where $k$ is the rank of a meaning within the set ${\cal M}_i$ when ordered by the frequency of the corresponding words. Thus meanings of high-frequency words are high-frequency confounders. Learning times from Monte Carlo simulations are shown in Fig.~\ref{RZL}.

We observe two distinct learning-time regimes.  At small $C$, the learning time is constant, and close to the time  it takes for all words in the lexicon to appear at least once. (This time is given by Eq.~(\ref{tast1}) with $a^\ast=0$). In this regime, learning is as fast as it can possibly be: mutual exclusivity is \emph{maximally efficient} and reverses the undesirable increase in learning times that arises from nonuniform confounder distributions. Above a critical context size, the learning time rises, but remains much smaller than when words are learned independently: mutual exclusivity is \emph{partially efficient} in this regime.

Our exact result (\ref{lme}) can be used to explain these observations, details of which appear in Appendix~\ref{app:b}. For the RZ model as described above, it turns out that only one confounder loop $\ell=(1,2)$ is relevant at late times. Consequently, the learning probability $L(t)$ is asymptotically given as the product of two factors. The first gives the probability that all words have been encountered by time $t$, and approaches unity exponentially with rate $1/\mu W$. The second is the probability that the loop $\ell=(1,2)$ has not decayed away: this approaches unity with rate $3(1-a^\ast)/2\mu$.  The appearance frequency of the most frequent confounder, $a^\ast$, increases with context size. When $a^\ast < 1-\frac{2}{3W}$, the slowest relaxational mode of the learning probability is associated with each word being uttered at least once, whereas for larger values, the slowest mode comes from eliminating the confounder loop. In this latter partially-efficient regime, the lexicon learning time is predicted as $t^\ast = - \frac{2\mu \ln\epsilon}{3(1-a^\ast)}$ for small $\epsilon$, in very good agreement with simulation data  (see Fig.~\ref{RZL}). We describe the sudden change in the dominant relaxational behavior---a phenomenon seen also in driven diffusive systems \cite{deGier05}---as a \emph{dynamical phase transition}.   It is broadly reminiscent of transitions exhibited by combinatorial optimization problems, whereby the number of unsatisfied constraints increases from zero above a critical difficulty threshold \cite{Mezard02}. In the present case the learning problem remains solvable in both regimes, but there is a transition from a regime where it is solved in constant time to one where the time grows super-exponentially in the difficulty of the problem (here, the context size).

%: Discussion

To summarize, we have found that mutual exclusivity is an extremely powerful word-learning heuristic. It can yield lexicon learning times in the presence of uncertainty that coincide with the time taken for each word to be heard at least once. Empirical data (summarized in \cite{BSS10}) suggests that this is easily fast enough for realistic lexicons of $W=60,000$ words to be learned. To enter the partially-efficient regime, each word's most frequent confounder would need to be present in at least $99.99\%$ of all episodes: even then, learning is over $W$ times faster than when mutual exclusivity is not applied. The dynamical transition between a maximally- and partially-efficient regime also appears to be present in a variety of word-learning models we have investigated, e.g., those in which confounder frequencies are uncorrelated with their corresponding word frequencies, or using less memory-intensive learning strategies \cite{RSSBip}. We also expect the transition to be evident in models where the target meaning does not always appear, at least in the regime where learning is possible \cite{Tilles12a,Tilles12b}.  We believe the analytical methods introduced in this work should allow more detailed quantities to be calculated, e.g., the distribution of learning times for a given word, which would shed light on such phenomena as the childhood vocabulary explosion at around 18 months \cite{McMurray07}.  Similar thinking may also allow analysis of other nonequilibrium dynamical systems whose master equations are hard to solve directly. Finally, our results suggest new empirical questions, such as whether high-frequency confounders correlate with high-frequency words, and the extent to which learners are able to apply the mutual-exclusivity constraint retroactively. We therefore contend that statistical physicists can contribute much to the understanding of how children learn the meaning of words.

\textit{Acknowledgments} --- We thank Mike Cates and Cait MacPhee for comments on the manuscript.

\appendix

\section{Learning time in the noninteracting case}
\label{app:a}

In the main text, we derived a formula---given there as Eq.~(1)---for the probability $L(t)$ that a lexicon of words is learned by time $t$ if they are learned independently by cross-situational learning.  This read
\begin{equation}
L(t) = \Big\langle \prod_{i} E_i(t) \prod_{j\ne i} [1 - A_{i,j}(t)] \Big\rangle
\end{equation}
where $E_i(t) = 1$ only if word $i$ has been presented by time $t$, and $A_{i,j}(t)$ is $1$ if word $i$ has never been presented, or, if in every presentation up to time $t$, the confounding meaning $j\ne i$ has always appeared alongside. The angle brackets denote an average over all possible exposure sequences.  Under all other conditions, these indicator variables are zero.

This equation was first of all presented in an alternative form which follows from the fact that $E_i(t)=0$ implies that $A_{i,j}(t)=1$ for all $j \ne i$.  Hence, for all allowed combinations of $E_i(t)$ and $A_{i,j}(t)$, we have the identity $[1-E_i(t)] A_{i,j}(t) = [1-E_i(t)]$ which can be rearranged to obtain $E_i(t)[1-A_{i,j}(t)] = [1-A_{i,j}(t)]$.  Assuming that there is at least one confounding meaning for each word, the $E_i(t)$ variables in the above equation are then redundant, and the more concise form
\begin{equation}
L(t) = \Big\langle \prod_i \prod_{j\ne i} [1 - A_{i,j}(t)] \Big\rangle
\end{equation}
then applies.

In the main text, we discussed models where contexts of confounding meanings were independently sampled from distributions that may be word-dependent, but remain fixed over time.  In particular, this implies that the contexts appearing against different words are independent, and we have factorization of the average into word-dependent factors:
\begin{equation}
\label{factor}
L(t) = \prod_i \Big\langle \prod_{j\ne i} [1 - A_{i,j}(t)] \Big\rangle \;.
\end{equation}
Since the confounder distributions are fixed, we find after $n_i$ presentations of word $i$ that
\begin{equation}
 A_{i,j_1}   \cdots A_{i,j_k} = \left\{ \begin{array}{ll}
1 & \mbox{with prob. $a_i(j_1, \ldots, j_k)^{n_i}$} \\
0 & \mbox{otherwise} \end{array} \right.
\end{equation}
where $a_i(j_1, \ldots, j_k)$ is the joint probability that all $k$ meanings $j_1, j_2, j_3, \ldots, j_k$ appear in a single episode.  Since word $i$ is presented as a Poisson process with frequency $\phi_i$, we find that
\begin{align}
\big\langle A_{i,j_1} A_{i,j_2}  \cdots A_{i,j_k} \big\rangle &= \sum_{n_i=0}^{\infty} \frac{(\phi_i t)^{n_i}}{n_i!} a_i(j_1, \ldots, j_k)^{n_i} {\rm e}^{-\phi_i t} \nonumber\\
&= {\rm e}^{-\phi_i [1 - a_i(j_1, \ldots, j_k)] t}
\end{align}
Therefore, on multiplying out the average in (\ref{factor}), we find a sum of exponential decays.  We are interested in the slowest decay mode, which corresponds to the highest possible value of $a_i(j_1, \ldots, j_k)$ among all possible sets of confounding meanings.  As noted in the main text, any combination of meanings $j_1, j_2, \ldots, j_k$ cannot appear more frequently than the least frequent meaning among that subset. If, for each word, the individual meaning frequencies $a_i(j)$ are distinct for different $j$, there will be a unique maximum appearance frequency, and the slowest decay is given by $a_i^\ast = \max_j \{ a_i(j) \}$. Multiplying the factors for each word $i$ together yields Eq.~(2) of the main text.  We note that in the special case where the most frequent meaning is $r$-fold degenerate, we acquire a prefactor $r$ in front of the dominant exponential decay.

For the case where $a_i^\ast$ is the same for all words $i$, Eq.~(3) in the main text is obtained by taking the logarithm of $L(t)$, replacing the sum with an integral, and expanding the logarithm to first order. For a Zipf distribution of word frequencies, $\phi_i = 1/(\mu i)$, $\mu = \sum_{i=1}^{W} (1/i)$, this procedure yields \cite{BSS10}
\begin{align}
\label{lgzipf}
\ln L(t) &\approx - \int_1^{W} {\rm d} x \exp\left( - \frac{(1-a^\ast) t}{\mu x} \right) \nonumber
\\ &\approx -\frac{\mu W^2 }{(1-a^\ast) t} \exp\left( -\frac{(1-a^\ast) t}{\mu W} \right) \;,
\end{align}
where we have used the asymptotics of the exponential integral \cite{AS65} to obtain the second approximate equality.  The solution of the equation $\ln L(t^\ast) = \ln(1-\epsilon)$ yields the learning time given by Eq.~(3).  This involves the Lambert W function which is defined by solutions of the equation $ {\cal W}(z){\rm e}^{{\cal W}(z)} = z$ \cite{Corless96}.

\section{Learning time for the interacting RZ model}
\label{app:b}

In the interacting RZ model described in the main text, it is assumed that all meanings are labeled and so Eq.~(4) for the learning probability simplifies to
\begin{equation}
L(t) = \left\langle \prod_{i=1}^{W} E_i(t) \prod_{\ell} \left[ 1 - C_{\ell}(t) \right] \right\rangle \;,
\end{equation}
where here $C_{\ell}(t) = A_{i_1,i_2}(t) A_{i_2,i_3}(t) \cdots A_{i_n,i_1}(t)$ for an ordered subset $\ell=(i_1, i_2, \ldots, i_n)$ of the $W$ meanings.

Numerical investigations of the RZ sampling procedure reveal that, when the context size is large, the most frequent confounders are all very likely to appear ($a_{i,j} \approx 1$ for the lowest $j$), while the relative frequencies of \emph{non-appearance} diverge with $C$, i.e., that $(1-a_{i,j+1})/(1-a_{i,j}) \to \infty$ as $C$ is increased.  Since it is these non-appearance probabilities, $1-a_{i,j}$, that enter into the decay rates of correlation functions (see above), it follows that the slowest-decaying confounder loops are those that are (a) short; and (b) comprise only the most frequent meanings.  The slowest decay of all loops therefore comes from $\ell=(1,2)$.  We have found that a good match between theory and numerical data is obtained by assuming this is the \emph{only} loop that contributes to the late-time behavior of $L(t)$.

To obtain the theoretical prediction, we first make this single-loop approximation:
\begin{align}
\label{Eaa}
L(t) &= \left\langle \prod_{i=1}^{W} E_i(t) \left[ 1 - A_{1,2}(t) A_{2,1}(t) \right] \right\rangle \nonumber\\
&= \prod_{i=1}^{W} \langle E_i(t) \rangle \left[ 1 - \frac{ \langle E_1(t)A_{1,2}(t) \rangle \langle E_2(t)A_{2,1}(t) \rangle }{ \langle E_1(t) \rangle \langle E_2(t) \rangle } \right]\;,
\end{align}
where we have used the fact that the contexts presented alongside different words are uncorrelated. $\langle E_i(t) \rangle$ is the probability that an event governed by a Poisson process with frequency $\phi_i$ has occurred at least once by time $t$.  Hence,
\begin{equation}
\prod_{i=1}^{W} \langle E_i(t) \rangle = \prod_{i=1}^{W} \left[1 - {\rm e}^{-\phi_i t}\right] \;.
\end{equation}
This is of the same form as Eq.~(3) of the main text, but with $a^\ast=0$, and so from (\ref{lgzipf}) we have that
\begin{equation}
\label{fm}
\prod_{i=1}^{W} \langle E_i(t) \rangle \approx \exp\left( - \frac{\mu W^2}{t} {\rm e}^{-t/\mu W} \right) \;.
\end{equation}

Turning now to the second term in (\ref{Eaa}), we use again the identity $[1-E_i(t)] A_{i,j}(t) = [1-E_i(t)]$ from the previous section, to find that $E_i(t) A_{i,j}(t) =  1 - A_{i,j}(t) - E_i(t)$.  Hence,
\begin{equation}
\Lambda_{i,j} = \frac{\langle E_i(t) A_{i,j}(t) \rangle}{\langle E_i(t) \rangle} = \frac{{\rm e}^{-\phi_i[1-a_{i}(j)]t} - {\rm e}^{-\phi_i t}}{1 - {\rm e}^{-\phi_i t}} \;.
\end{equation}
If $a_{i}(j)$ is close to unity, as is the case for the high-frequency meanings in the RZ model, we have at late times that
\begin{equation}
\Lambda_{i,j} \sim {\rm e}^{-\phi_i[1-a_{i}(j)]t} \;.
\end{equation}

Combining this result with (\ref{fm}) in (\ref{Eaa}), and noting that $\phi_i = 1/(\mu i)$, we arrive at 
\begin{equation}
\label{Ltpt}
L(t) \sim \exp\left( - \frac{\mu W^2 {\rm e}^{-\frac{t}{\mu W}} }{t} \right) \left( 1 - \exp\left[ -\frac{3(1-a^\ast)t}{2\mu} \right] \right) 
\end{equation}
which gives an asymptotic expression for the learning probability as a function of time.  To convert this into a learning time, we need to solve the equation $L(t^\ast)=1-\epsilon$.  Unfortunately, we have not been able to do this exactly.  It is however straightforward now to identify the slowest decay mode of $L(t)$ by expanding out:
\begin{equation}
\label{Lexp}
L(t) \approx 1 - \frac{\mu W^2}{t} \exp\left({-\frac{t}{\mu W}}\right) - \exp\left(-\frac{3(1-a^\ast)t}{2\mu}\right) + \cdots \;.
\end{equation}
Thus, as stated in the main text, we find that the mode associated with exposure of the entire lexicon decays at a rate $1/\mu W$, and that the mode associated with elimination of the confounder loop decays at rate $3(1-a^\ast)t / 2 \mu$.

Now, as $\epsilon\to0$, the learning time $t$ must diverge towards infinity.  Hence, as $\epsilon$ is reduced, the subleading term in (\ref{Lexp}) can be made arbitrarily small, and the solution of
\begin{equation}
\epsilon = \left\{ \begin{array}{ll} 
\frac{\mu W^2}{t^\ast} \exp\left({-\frac{t^\ast}{\mu W}}\right) & \mbox{for $a^\ast < 1-\frac{2}{3W}$} \\
\exp\left(-\frac{3(1-a^\ast)t^\ast}{2\mu}\right) & \mbox{for $a^\ast > 1-\frac{2}{3W}$}
\end{array}\right.
\end{equation}
yields the learning time $t^\ast$ to better and better accuracy in the limit $\epsilon\to0$.  Formally, as $\epsilon\to0$, the function $\phi(a^\ast) = t^\ast / \ln\epsilon$ exhibits a nonanalyticity at $a^\ast=1-\frac{2}{3W}$. It is in this sense that we regard this model to exhibit a dynamical phase transition.

For more general models, in which more than one candidate loop enters at large times, we have found that including only loops of length 2 in the product in (\ref{lme}) yields very good agreement with simulation data. More precisely, numerically-determined roots of $L(t) = 1-\epsilon$ with $L(t)$ given by the approximate expression
\begin{equation}
L(t) \approx \prod_{i=1}^{W} \left[ 1 - {\rm e}^{-\phi_i t} \right] \prod_{\langle i,j \rangle} \left[ 1 - \Lambda_{i,j} \Lambda_{j,i} \right]
\end{equation}
correspond well with simulation data, and furthermore provides evidence for our claim that the dynamical phase transition reported is not a peculiarity of the specific model discussed in the main text.

\end{document}